\documentclass[aps,prl,floatfix,twocolumn,reprint]{revtex4}
\usepackage{eurosym}
\usepackage{amsbsy}
\usepackage{latexsym,epsfig,graphicx}
\usepackage{dcolumn}
\usepackage{graphicx}
\usepackage{subfigure}
\usepackage{comment}
\usepackage{color}
\usepackage{bm}
\usepackage{mathrsfs}
\usepackage{amssymb}
\usepackage{amsfonts}
\usepackage{mathrsfs}
\usepackage{amsmath}
\usepackage{color}
\usepackage{graphicx}
\usepackage{bm}
\usepackage{amssymb}
\usepackage{xspace}
\usepackage{epstopdf}
\usepackage{dcolumn}
\usepackage{tabularx}
\usepackage{longtable}
\usepackage{diagbox}
\usepackage[colorlinks=true, letterpaper=true, pdfstartview=FitV, linkcolor=blue, citecolor=blue, urlcolor=blue]{hyperref}
\usepackage[normalem]{ulem}
\usepackage[version=4]{mhchem}

\pdfoutput=1

\begin{document}

\title{Topological Triply Degenerate Points Induced by Spin-Tensor-Momentum Couplings}
\author{Haiping Hu}
\affiliation{Department of Physics, The University of Texas at Dallas, Richardson, Texas
75080, USA}
\author{Junpeng Hou}
\affiliation{Department of Physics, The University of Texas at Dallas, Richardson, Texas
75080, USA}
\author{Fan Zhang}
\affiliation{Department of Physics, The University of Texas at Dallas, Richardson, Texas
75080, USA}
\author{Chuanwei Zhang}
\email{chuanwei.zhang@utdallas.edu}
\affiliation{Department of Physics, The University of Texas at Dallas, Richardson, Texas
75080, USA}

\begin{abstract}
The recent discovery of triply degenerate points (TDPs) in topological
materials has opened a new perspective toward the realization of novel
quasiparticles without counterparts in quantum field theory. The emergence
of such protected nodes is often attributed to spin-vector-momentum
couplings. Here we show that the interplay between spin-tensor- and
spin-vector-momentum couplings can induce three types of TDPs, classified by
different monopole charges ($\mathcal{C}=\pm 2,\pm 1,0$). 
A Zeeman field can lift them into Weyl points with distinct numbers and
charges. Different TDPs of the same
type are connected by intriguing Fermi arcs at surfaces, and transitions
between different types are accompanied by level crossings along
high-symmetry lines. We further propose an experimental scheme to realize
such TDPs in cold-atom optical lattices. Our results provide a framework for
studying spin-tensor-momentum coupling-induced TDPs and other exotic
quasiparticles.
\end{abstract}

\maketitle

{\color{blue}\textit{Introduction.---}}Topological states of matter~\cite%
{review1,review2} provide a fertile ground for discovering new
quasiparticles in condensed matter physics, such as Weyl~\cite%
{weylbook,weylmurakami,weylxgwan,weylbalents,weyl1,weyl2,weyl3,weyl4,weyl5,weyl6,weyl7,weyl8,weyl9,weyl10,CCL,ortix,ypz}
and Dirac fermions~\cite{ypz,dirac1,dirac2,dirac3,dirac4,dirac5,dirac6} that
were originally predicted in high-energy physics and recently observed in
solid-state materials~\cite{semimetal}. In topological semimetals, Weyl and
Dirac points correspond to two- and four-fold degenerate linear band
crossing points, hallmarks of relativistic particles with half-integer
spins. Remarkably, the recent discovery of triply-degenerate points (TDPs)~%
\cite{tf1,tf2,tf3,tf4,tf5,tf6,tf7,tf8,tf9,tf10,tf11,tf12,tf13} in
semimetals has opened an avenue for exploring new types of quasiparticles
that have no analog in quantum field theory. Such TDPs possess effective
integer spins while preserving Fermi statistics and linear dispersions.

Generally, the linearly dispersed quasiparticles near band degeneracies can
be described by Hamiltonians with a spin-vector-momentum coupling $\sim%
\bm{k}\cdot\bm{F}$, where $\bm{F}=(F_x,F_y,F_z)$ is a spin-vector. A
degenerate point acts like a magnetic monopole in momentum space with a
topological charge $\mathcal{C}$ determined by the quantized Berry flux
emanating from the point. In this context, a TDP with $F\!=\!1$ behaves like
a three-component fermion with $\mathcal{C}\!=\!\pm 2$. However, it is well
known that a full description of any large spin with $F\!\geq\!1$ naturally
involves spin-tensors up to rank $2F$. For instance, there exist six rank-$2$
spin quadrupole tensors $N_{ij}=(F_{i}F_{j}+F_{j}F_{i})/2-\delta _{ij}\bm{F}%
^{2}/3$ for $F\!=\!1$ in addition to the three vector components $F_i$ ($%
i\!=\!x,y,z$). Therefore two questions naturally arise. Can
spin-tensor-momentum couplings produce novel types of TDPs with distinct
topological properties? If so, how can such novel TDPs and associated
spin-momentum couplings be realized in realistic systems?

In this paper, we address these two important questions by showing that two
novel types of TDPs can emerge from the interplay between spin-vector- and
spin-tensor-momentum couplings, and cold-atom optical lattices provide an
attractive platform for their realizations. We call the TDPs described by
the spin-vector-momentum coupling type-I~\cite%
{tf1,tf2,tf3,tf4,tf5,tf6,tf7,tf8,tf9,tf10,tf11,tf12,tf13} and the TDPs
induced by spin-tensor-momentum couplings types II and III. Here are our main
results. First, the three types have different topological charges: $%
\mathcal{C}=\pm 2$, $\pm 1$, and $0$ for types I, II, and III, respectively.
A Zeeman field can lift them into Weyl points with distinct numbers and
charges. 

Second, the topological transitions between different types, accompanied by
level crossings along high-symmetry lines, can be achieved by tuning the
relative strengths of spin-vector- and spin-tensor-momentum couplings. By
constructing a minimum three-band lattice model, we display different types
of TDPs in the bulk and their exotic Fermi arcs at the surface.

Thirdly, since the type-II and type-III TDPs have not been discovered before, we
propose the first experimental scheme for realizing type-II and required
spin-momentum couplings using cold atoms in an optical lattice.
Spin-vector-momentum coupling is crucial for many important condensed matter
phenomena, and its recent experimental realization in ultracold atomic 
gases~\cite{1dsoc1,1dsoc2,1dsoc3,1dsoc4,1dsoc5,1dsoc6,1dsoc7,soc2,soc3,1dsoc8,soc1,spintensor1}%
has provided a highly controllable and disorder-free platform for exploring
topological quantum matter. In cold atoms, spins are modeled by atomic
hyperfine states, and a spin with $F\!\geq \!1$ can be naturally obtained.
Nowadays, various types of spin-vector-momentum coupling for both spin-$1/2$
and spin-$1$ have been proposed and realized~\cite%
{1dsoc1,1dsoc2,1dsoc3,1dsoc4,1dsoc5,1dsoc6,1dsoc7,soc2,soc3,1dsoc8,soc1,3dsoc1,lxj,spintensor1,Wu2016}%
. A scheme for realizing spin-tensor-momentum coupling of spin-$1$ atoms has
also been proposed recently~\cite{spintensor2} with ongoing experimental
efforts \cite{STMCexp}. Our scheme is built on these experimentally
available setups \cite{soc1,lxj} and may even pave the way for identifying
solid-state materials with our novel types of TDPs.

\begin{figure}[t]
\centering
\includegraphics[width=3.4in]{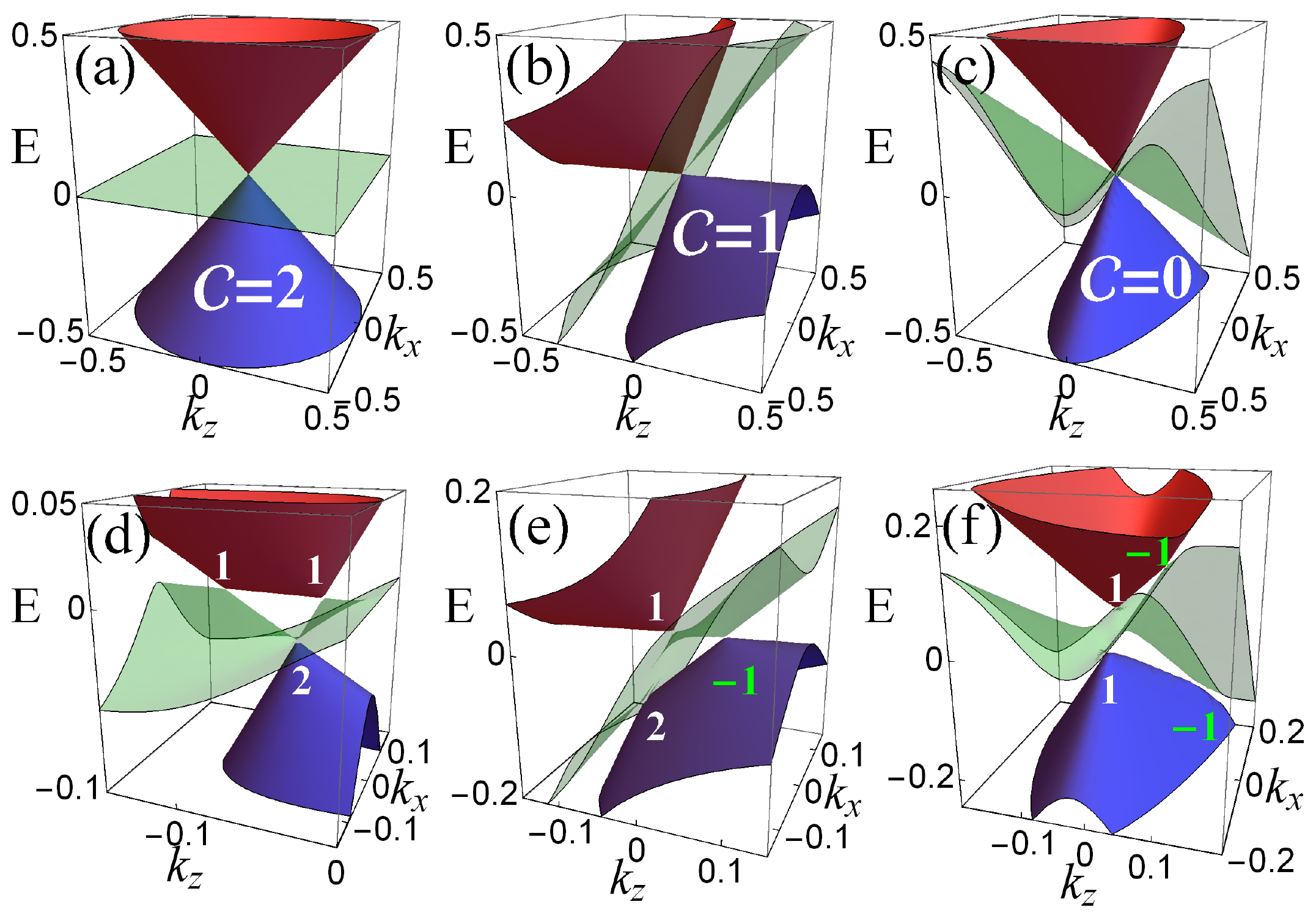}
\caption{(a)-(c) Band structures of three types of TDPs in the $k_{y}=0$ plane for model~(%
\protect\ref{model}). (a) The type-I with $\protect\alpha =1$ and $\protect\beta %
=0$. (b) The type-II with $\protect\alpha =1$, $\protect\beta =2$, 
and $N_{ij}$ is chosen as $F_{z}^{2}$. (c) The type-III with $\protect\alpha =1$, $\protect\beta =3$%
, and $N_{ij}=N_{xz}$. (d)-(f) Splittings of three types of TDPs due to a
Zeeman perturbation $\protect\varepsilon F_{z}$ with $\protect%
\varepsilon =0.05$. (d) The type-I splits into two linear Weyl points with $%
\mathcal{C}=1$ and one double-Weyl point~\cite{multiweyl} with $\mathcal{C}=2$; 
note that $\protect\beta =0.5$ instead of $0$ is used. (e)
The type-II splits into two linear Weyl points with $\mathcal{C}=\pm 1$ 
and one double-Weyl point~\cite{multiweyl} with $\mathcal{C}=2$. (f) The type-III splits into four linear
Weyl points with $\mathcal{C}=\pm 1$.}
\label{fig1}
\end{figure}

{\color{blue}\textit{Triply-degenerate points.---}}As a direct extension of
a two-fold degenerate Weyl point described by $H=\bm k\cdot \bm{\sigma }$,
the simplest TDP should be described by $H=\bm k\cdot \bm F$ with the spin-$%
1 $ vector $\bm F$~\cite%
{tf1,tf2,tf3,tf4,tf5,tf6,tf7,tf8,tf9,tf10,tf11,tf12,tf13}. The band
structure around such a TDP is shown in Fig.~\ref{fig1}(a), with a flat band located at the center and linear dispersions along all
directions for the three bands. We label the band indices $n$ for the
lower, middle, and upper bands as $-1$, $0$, and $1$, respectively. The corresponding wave function for band $n$ is denoted as $|\psi_n(\bm k)\rangle$.
The topological property of the TDP can be characterized by the first Chern numbers
\begin{equation}
\mathcal{C}_{n}=\frac{1}{2\pi }\oint_{\bm{S}}\bm{\Omega }_{n}(\bm{k})\cdot d%
\bm{S},  \label{topoinvariant}
\end{equation}%
where $\bm{S}$ is a closed surface enclosing the TDP 
and $\bm{\Omega }_{n}(\bm{k})\!=\!\bm{\nabla }_{\bm{k}}\times \langle
\psi _{n}(\bm{k})|i\bm{\nabla }_{\bm{k}}|\psi _{n}(\bm{k})\rangle $ is the
Berry curvature of band $n$. For $H\!=\!\bm{k}\cdot \bm{F}$, $\bm{\Omega }%
_{n}(\bm{k})\!=\!-n\bm{k/}k^{3}$, yielding $\mathcal{C}_{n}\!=\!-2n$ for
the three bands. The monopole charge $\mathcal{C}$ can be defined as the Chern
number of the lower band, i.e., $\mathcal{C}\!=\!\mathcal{C}_{-1}$. 
Thus, this simplest TDP has $\mathcal{C}=2$ and behaves as a
momentum-space monopole carrying two monopole charges.

Novel types of TDPs can emerge when spin-tensors are also considered. Since
a constant spin-tensor perturbation $\sim N_{ij}$ would break the three-fold 
degeneracy of $H=\bm{k}\cdot \bm{F}$, the stabilization of novel TDPs 
with linear dispersions requires the coupling of spin-tensors
with momentum. For general linear Hamiltonians with $H(\bm k)=-H(-\bm k)$, 
the property $\bm{\Omega}_{n}({\bm k})=\bm{\Omega} _{-n}(-\bm k)$ dictates 
$\mathcal{C}_{+1}\!=\!-\mathcal{C}_{-1}$ for the upper and lower bands and $%
\mathcal{C}_{0}\!=\!0$ for the middle one. Moreover, it can be proved that $%
\left\vert \mathcal{C}_{n}\right\vert \leq 2$ for such linear Hamiltonians
\cite{supplementary}. Therefore the monopole charges for TDPs can only be $%
\pm 2$, $\pm 1$, and $0$, indicating all possible TDPs can be classified
into three types: type-I with $\mathcal{C}=\pm 2$, type-II with $\mathcal{C}%
=\pm 1$, and type-III with $\mathcal{C}=0$.

All three types of TDPs can be illustrated using the following simple model~\cite%
{supplementary}:
\begin{equation}
H({\bm{k}})=k_{x}F_{x}+k_{y}F_{y}+k_{z}(\alpha F_{z}+\beta N_{ij}),
\label{model}
\end{equation}%
where the spin-tensor $N_{ij}$ is coupled to $k_z$. 
By choosing different types of spin-tensors and tuning the relative strength of
spin-tensor-momentum coupling $\beta/\alpha$, we find that
(i) the three spin-tensors $N_{xx}$, $N_{yy}$, and $N_{xy}$ do not change the 
monopole charge $\mathcal{C}\!=\!\pm 2$ of the type-I TDP,
(ii) the tensor $N_{zz}$ induces a $\mathcal{C}\!=\!\pm 1$ TDP for 
$\left\vert \beta \right\vert >\left\vert\alpha \right\vert \neq 0$, 
dubbed type-II and depicted in Fig.~\ref{fig1}(b),
and (iii) the tensor $N_{xz}$ or $N_{yz}$ induces a $\mathcal{C}\!=\!0$ TDP for 
$\left\vert \beta \right\vert >2\left\vert \alpha\right\vert \neq 0$, 
dubbed type-III and depicted in Fig.~\ref{fig1}(c). 
Markedly, the energy dispersions are linear around all these three types of TDPs. 

\begin{figure}[t]
\centering
\includegraphics[width=3.4in]{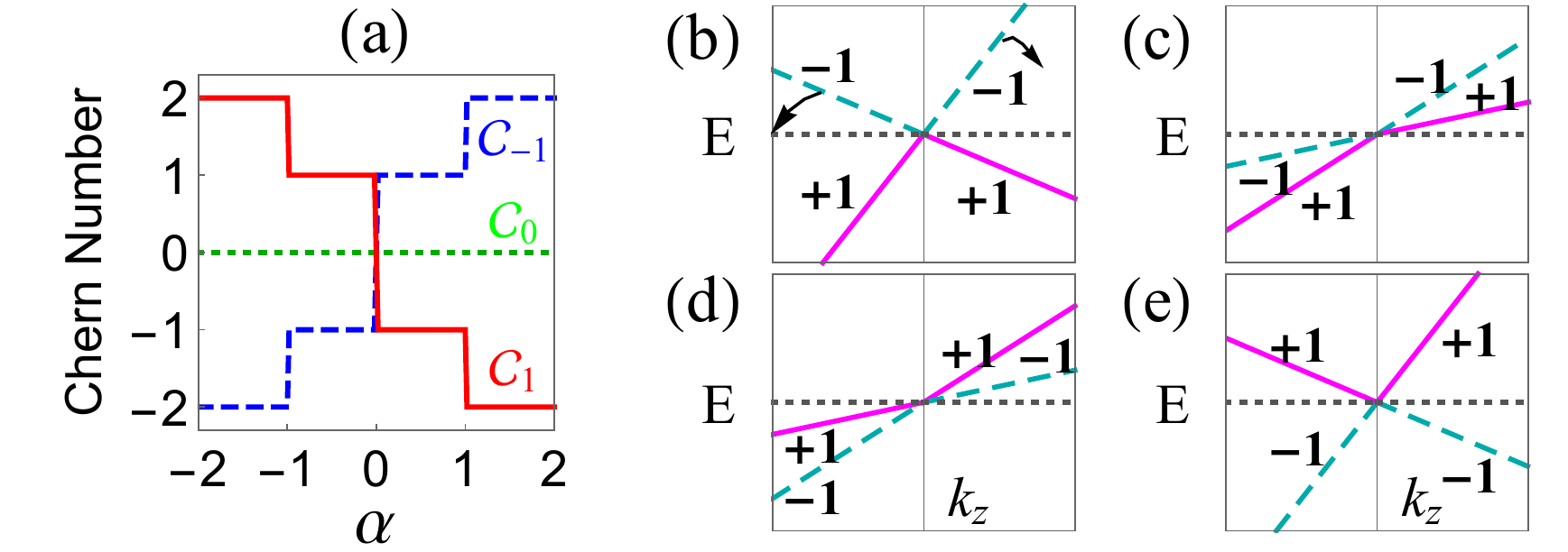}
\caption{Phase transitions between type-I and type-II TDPs by tuning $\protect%
\alpha$ while fixing $\protect\beta =1$ in Eq.~(\protect\ref{model}). (a)
Chern numbers as functions of $\protect\alpha$ for the lower (dashed blue),
middle (dotted green), and upper (solid red) bands. (b)-(e) Band structures
along the $k_{x}=k_{y}=0$ line with $\protect\alpha\!=\!2$,~$0.5$,~$-0.5$,
and~$-2$, respectively. We have labeled the Chern contributions of each branch: $+1$
(solid magenta), $-1$ (dashed cyan), and $0$ (dotted black).}
\label{fig2}
\end{figure}

{\color{blue}\textit{Type-II TDPs.---}}Type-II TDPs can be induced from
the type-I by choosing $N_{ij}$ as $F_{z}^{2}=N_{zz}+\frac{2}{3}$ in Eq.~(\ref%
{model}). Since the additional spin-independent term $2\beta k_{z}/3$ does
not affect the eigenstates or any topological transition, 
we use $F_{z}^{2}$ instead of $N_{zz}$ for better presentation of our results. 
To study the transition between
type-I and type-II TDPs due to the competition between spin-vector- and
spin-tensor-momentum couplings, we fix $\beta =1$, vary $\alpha $, and
calculate $\mathcal{C}_{n}$ numerically using Eq.~(\ref{topoinvariant}). As
exhibited in Fig.~\ref{fig2}(a), the lower-band Chern number $\mathcal{C}%
_{-1}$ (the monopole charge) changes from $2$ (type-I) to $\pm 1$
(type-II), and then to $-2$ (type-I) with decreasing $\alpha$.

The topological transitions can be understood by the band crossings \cite%
{supplementary} along the $k_{x}=k_{y}=0$ line, as sketched in Figs.~\ref%
{fig2}(b)-\ref{fig2}(e). Note that the Chern number of each band has two
contributions from the $k_{z}<0$ and $k_{z}>0$ branches in the surface
integral of Eq.~(\ref{topoinvariant}): $\mathcal{C}=\mathcal{C}_{k_{z}<0}+%
\mathcal{C}_{k_{z}>0}$. When $\alpha >1$, the spin-vector-momentum coupling 
$k_{z}F_{z}$ dominates and the model~(\ref{model}) is adiabatically
connected to $H=\bm{k}\cdot \bm{F}$ (type-I with $\mathcal{C}=2$); the
contributions from the two branches of the lower band are $\mathcal{C}%
_{k_{z}<0}=\mathcal{C}_{k_{z}>0}=+1$, as shown in Fig.~\ref{fig2}(b). 
With decreasing $\alpha $, the $k_{z}<0$ ($k_{z}>0$) branch of the lower band rotates
clockwise (counterclockwise) in the $E$-$k_{z}$ plane. At $\alpha =1$, the
middle band crosses simultaneously with the $k_{z}<0$ branch of the upper
band and $k_{z}>0$ branch of the lower band. After the band crossing, as
shown in Fig.~\ref{fig2}(c), the lower band consists of two branches with
Chern contributions $\mathcal{C}_{k_{z}<0}=1$ and $\mathcal{C}_{k_{z}>0}=0$,
yielding a type-II TDP with $\mathcal{C}=1$, in consistent with numerical
results. With further decreasing $\alpha$, another level crossing occurs
between the middle band and the $k_{z}<0$ ($k_{z}>0$) branch of lower
(upper) band at $\alpha =0$, as shown in Fig.~\ref{fig2}(d). This crossing
changes $\mathcal{C}$ from $1$ to $-1$ and the resulting TDP is still
type-II. A third band crossing occurs at $\alpha =-1$. For $\alpha <-1$, all
bands are totally reversed compared to the $\alpha >1$ case as shown in Fig.~%
\ref{fig2}(e), and the TDP is of type-I with $\mathcal{C}=-2$.

Type-I and type-II TDPs can be broken into different two-fold degenerate
Weyl points in the presence of perturbations. With an additional
Zeeman term $\varepsilon F_{z}$ ($\varepsilon \ll 1$) to Eq.~(\ref{model}), 
the eigenspectrum of the total Hamiltonian shows that
both types of TDPs are broken into three nodal points located at $W_{\pm
}=(0,0,-\varepsilon /(\alpha \pm \beta ))$ and $W_{3}=(0,0,-\varepsilon
/\alpha )$ \cite{supplementary}, as illustrated in Figs.~\ref{fig1}(d)-\ref{fig1}%
(e). The first two at $W_{\pm }$ are linear Weyl points, which have the same
charge $\mathcal{C}\!=\!1$ for type-I ($|\beta |<|\alpha |$) but opposite
charges $\mathcal{C}\!=\!\pm 1$ for type-II ($|\beta |>|\alpha |$)~\cite%
{supplementary}. The third node at $W_{3}$ is a multi-Weyl point \cite%
{multiweyl} with $\mathcal{C}=2$, whose dispersion is linear in the $k_{z}$
direction but quadratic along the other two directions due to the indirect
couplings between the lower and upper bands by $F_{x}$ and $F_{y}$.

Splittings of TDPs can be understood using Fig.~\ref{fig2} with the small
Zeeman field effectively lifting the middle band. For type-I in Fig.~\ref%
{fig2}(b), the horizontal band would cross the two branches with the same
Chern contributions, resulting in two linear Weyl points of the same
monopole charge. Apart from the two linear Weyl points, there still exists a
two-fold degenerate point with $\mathcal{C}=2$. By contrast, type-II in Fig.~%
\ref{fig2}(c) has a different configuration of energy levels, and the
horizontal band would cross the two branches with opposite Chern contributions,
leading to two linear Weyl points carrying opposite charges.

\begin{figure}[t]
\centering
\includegraphics[width=3.3in]{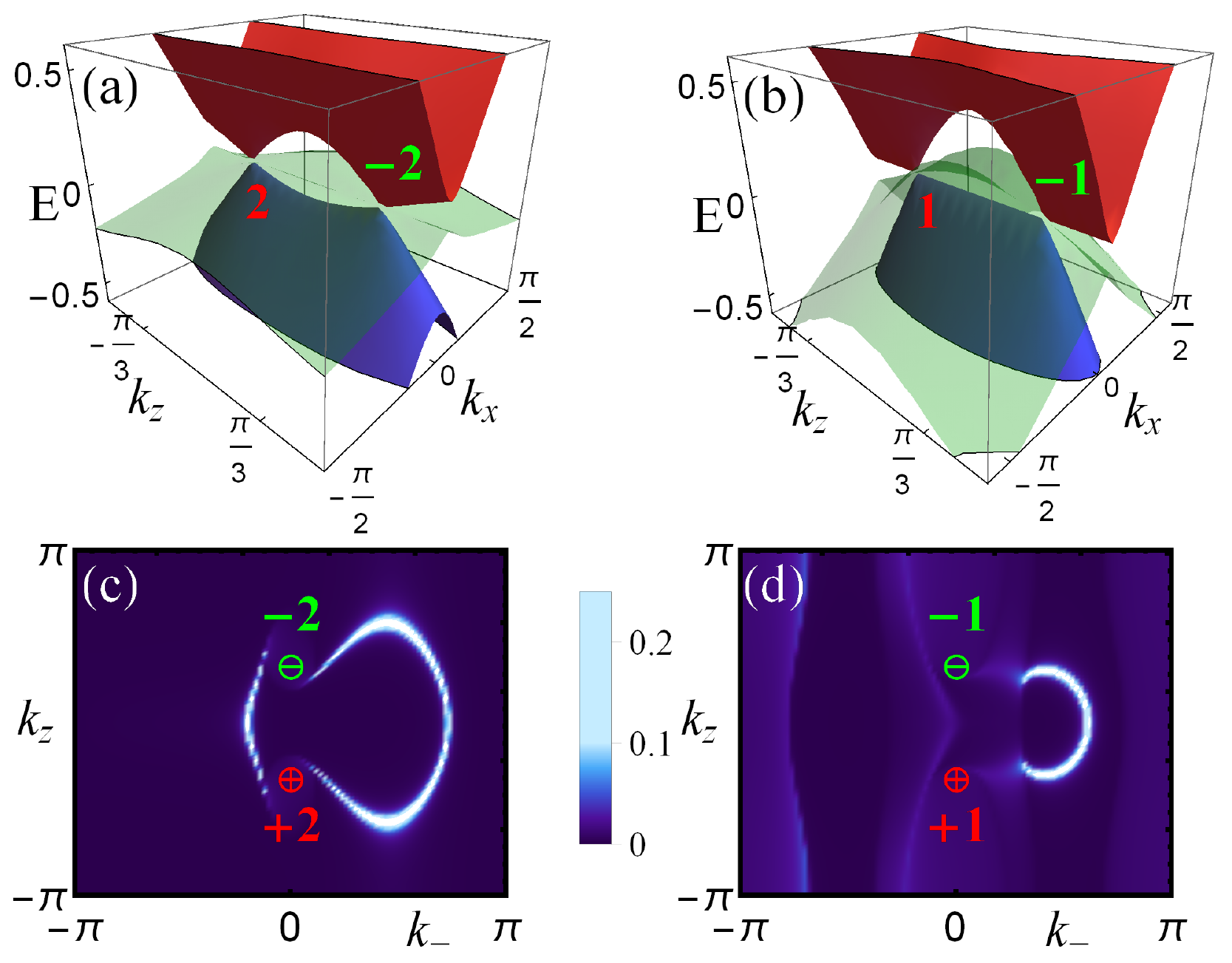}
\caption{Bulk band structures with TDPs and $(110)$ surface spectral
densities with Fermi arcs of model~(\protect\ref{toymodel}). 
(a) and (c) the type-I with $\mathcal{C}=\pm 2$ and two surface arcs. 
(b) and (d) the type-II with $\mathcal{C}=\pm 1$ and only one surface arc. In both
cases, two TDPs appear at $(0,0,\pm \protect\pi /3)$, and each projected
node is marked by its monopole charge. In our calculation, $\protect\gamma%
=-0.5$, $\protect\omega=0.25$, and $t_{0}=0.5$ are used; $\protect\beta =0.5$
is used in (a) and (c) while $\protect\beta =1.5$ in (b) and (d).}
\label{fig3}
\end{figure}

{\color{blue}\textit{Surface Fermi arcs.---}}For a 3D Weyl semimetal, it is
well known that a Fermi arc exists in the 2D surface Brillouin zone
connecting two projected Weyl points of opposite charges~\cite{weylxgwan}.
In the above discussions, we have seen that there exist TDPs of opposite
charges for both type-I and type-II. Therefore, it is important to examine
and compare their surface consequences. The coexistence of TDPs with
opposite charges can be best illustrated by the following minimal model on a cubic lattice:
\begin{eqnarray}
H(\bm{k}) &=&F_{x}\sin k_{x}+F_{y}\sin k_{y}+t_{0}(F_{z}+\beta F_{z}^{2})
\notag \\
&&(\cos k_{x}+\cos k_{y}+\cos k_{z}-2+\gamma ),  \label{toymodel}
\end{eqnarray}%
which hosts two TDPs at $\bm{k}\!=\!(0,0,\pm\arccos(-\gamma ))$ for $|\gamma|<1$. 
As displayed in Figs.~\ref{fig3}(a)-\ref{fig3}(b), the band structure of
model~({\ref{toymodel}}) with $\gamma \!=\!-0.5$ features two TDPs at $%
(0,0,\pm \pi /3)$. Around the two TDPs, the Hamiltonians can be expanded as $%
H_{\pm }(\delta \bm{k})=\delta k_{x}F_{x}+\delta k_{y}F_{y}\mp \frac{\sqrt{3}%
t_{0}}{2}\delta k_{z}(F_{z}+\beta F_{z}^{2})$ to the linear order. The above
effective Hamiltonian has the standard form of model~(\ref{model}), and 
the higher-order corrections would not affect the topological properties.
Therefore, the two TDPs belong to type-I for $|\beta|<1$ and type-II for $|\beta |>1$.

To reveal and compare the surface hallmarks of the two types of TDPs, we
impose a semi-infinite geometry with a $(110)$ surface in our calculation.
The surface Brillouin zone is expanded by $(k_{-},k_{z})$ with $%
k_{-}\!=\!\left( k_{x}-k_{y}\right) /\sqrt{2}$. Since the middle band
occupies most of the surface Brillouin zone at zero energy, we calculate the
surface spectral density $A(\omega ,\bm{k})\!=\!\text{Im}G(i\omega ,\bm{k}%
)/\pi $ \cite{tf1} at a finite $\omega $ in order to distinguish the surface
and bulk states. Here $G\!=\!(i\omega -H)^{-1}$ is the single-particle
Green's function. For type-I, there is a pair of Fermi arcs, 
and each emanates from one projected TDP and ends at the other, 
as illustrated in Fig.~\ref{fig3}(c). This clearly demonstrates
the double monopole charges of type-I TDPs. For type-II, the two projected
TDPs are connected by only one Fermi arc, as depicted in Fig.~\ref{fig3}(d).
This agrees well with the single monopole charges of type-II TDPs.

\begin{figure}[t]
\includegraphics[width=3.4in]{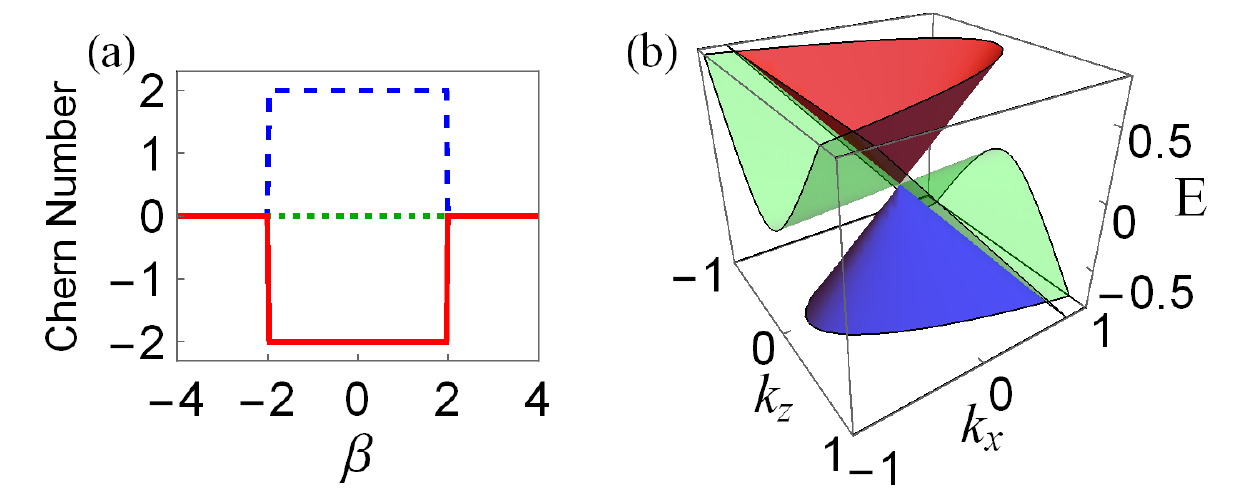}
\caption{Phase transitions between type-I and type-III TDPs by tuning $\protect%
\beta$ while fixing $\protect\alpha=1$ in Eq.~(\protect\ref{model}). (a)
Chern numbers as functions of $\protect\beta$ for the lower (dashed blue),
middle (dotted green), and upper (solid red) bands. (b) Band crossing at the
two transition lines with $\protect\beta =2$.}
\label{fig4}
\end{figure}

{\color{blue}\textit{Type-III TDPs.---}}Spin-tensor $N_{xz}$ or $N_{yz}$ in
model~(\ref{model}) can induce the topological transition of a TDP from type-I to
type-III. Here we use $N_{ij}=N_{xz}$ and $\alpha =1$ 
to illustrate the transition~\cite{supplementary} . The TDP is of
type-I for $|\beta |<2$ and type-III for $|\beta |>2$. At $|\beta |=2$, the
bands cross along two lines $k_{z}\pm k_{x}=k_{y}=0$, as shown in Fig.~\ref%
{fig4}(b). At one of these two line nodes, e.g., $k_{z}-k_{x}=k_{y}=0$, the
band energies are found to be $-\beta k_{z}/2$ and $(\beta \pm \sqrt{%
32+\beta ^{2}})k_{z}/4$. Clearly, at $\beta =2$ the
upper (lower) and middle bands cross at the $k_{z}<0$ ($k_{z}>0$) branch.
The band crossing of the other line node is rather similar. Because each
band crossing changes the Chern number by $1$, and the crossings along the
two lines are in the same branch, the Chern number must be changed by $2$ as
shown in Fig.~\ref{fig4}(a), yielding a transition of the TDP from type-I with $%
\mathcal{C}=\pm 2$ to type-III with $\mathcal{C}=0$.

Although a type-III TDP has vanishing Chern numbers, it can exhibit non-trivial
topological properties after breaking into linear Weyl points~\cite%
{supplementary} in the presence of a small Zeeman field, as depicted in Fig.~%
\ref{fig1}(f). There exist four Weyl points with $\mathcal{C}=\pm 1$ located
at $(k_{x},k_{z})=(\pm \beta \varepsilon /(\beta -2\alpha ),2\varepsilon
/(\beta -2\alpha ))$ and $(\pm \beta \varepsilon /(\beta +2\alpha
),-2\varepsilon /(\beta +2\alpha ))$ in the $k_{y}=0$ plane. From the above
discussions, we can see that the three types of TDPs have different patterns of Weyl
points after splitting, which results in distinct surface states since the
surface Fermi arcs can only connect two Weyl points of opposite charges.
Therefore, while for type-I the Fermi arcs only connect Weyl points
originating from different TDPs, for type-II and type-III there may exist
Fermi arcs connecting the Weyl points originating from the same TDP.
These features may be used to identify TDPs of different types.

{\color{blue}\textit{Experimental realization and observation.---}}The
type-II TDPs can be realized by coupling three atomic hyperfine states (e.g.,
the $6^{2}S_{1/2}$ ground-state manifold of $^{133}$Cs atom: $%
g_{1}\!=\!|4,-4\rangle $, $g_{0}\!=\!|3,-3\rangle $, and $%
g_{-1}\!=\!|4,-2\rangle $) using Raman beams in a spin-dependent square
lattice~\cite{supplementary}. The three states are used for mimicking the
spin-$1$ degree of freedom, and the proposed scheme is based on techniques
used in the recent experimental realization of 2D Rashba spin-orbit coupling
for spin-1/2 in optical lattices~\cite{soc1}. The atom-light interactions
include two crucial parts. One part is used for generating a spin-dependent
square lattice potentials $V_{g_{\pm 1}}\!\propto \![\sin (2k_{0}x)+\sin
(2k_{0}y)]$ and $V_{g_{0}}\!\propto \![-\sin (2k_{0}x)-\sin (2k_{0}y)]$ in
the $x$-$y$ plane by one laser beam~\cite{lxj}. In the tight-binding limit,
the $g_{+1}$ and $g_{-1}$ components stay on the same lattice sites. The
other part is used for inducing the required spin-momentum couplings between
the three hyperfine states, which can be achieved by adding another three
Raman beams $\bm{E}_{R_{1},R_{3}}=E_{R_{1},R_{3}}e^{\mp ik_{m}z}[\hat{{\bm{x}%
}}\cos (2k_{0}y)\mp \hat{{\bm{y}}}\cos (2k_{0}x)]$ and $\bm{E}%
_{R_{2}}=E_{R_{2}}e^{ik_{1}z}(i\hat{{\bm{x}}}+\hat{{\bm{y}}})$. The
resulting Raman couplings between $g_{\pm 1}$ and $g_{0}$ are $M_{\pm
1,0}\propto e^{i(k_{1}\pm k_{m})z}[\cos (2k_{0}x)\pm i\cos (2k_{0}y)]$ \cite%
{soc1,supplementary}. Because the spatially dependent phase factors contain
both spin-vector and spin-tensor components, they would produce both
spin-vector- and spin-tensor-momentum couplings $\sim {k_{z}}%
(k_{1}F_{z}^{2}+k_{m}F_{z})$ in a chosen gauge. A careful analysis of the
tight-binding model on the square lattice shows that the band
structure contains two TDPs located at $(0,0)$ and $(\pi ,0)$ in the $k_{x}$-%
$k_{y}$ plane at a constant $k_{z}$, similar to the case of model~(\ref%
{toymodel}). Around these TDPs, the effective Hamiltonians
have the standard form of Eq.~(\ref{model}), with the emergence of
spin-tensor-momentum coupling. Type-II TDPs require $|k_{1}|>|k_{m}|$, which
is naturally realized here since $k_{1}\approx \sqrt{k_{m}^{2}+4k_{0}^{2}}$
in our scheme~\cite{supplementary}.

The linear band dispersions and the three-fold degeneracy of a TDP may be
detected experimentally using the momentum-resolved radio-frequency spectroscopy
\cite{Jin2008}, as demonstrated in recent experiments for 2D spin-orbit
coupled atomic gases through spin-injection methods \cite%
{1dsoc6,1dsoc7,soc2,soc3}. Moreover, when the atomic gas is confined in a
hard wall box potential similar to those realized in recent experiments~\cite%
{box1,box2}, surface Fermi arcs would emerge at the boundary, which may
also be observed using the momentum-resolved radio-frequency spectroscopy.

{\color{blue}\textit{Discussions.}---}We have proposed and
demonstrated that the interplay between spin-vector- and
spin-tensor-momentum couplings can induce two novel types of TDPs possessing
distinct topological properties (e.g., Chern numbers, breaking into Weyl
points, surface Fermi arcs, etc.) from the already discovered type-I TDP in
solid-state materials. In particular, our proposed spin-tensor-momentum
coupling mechanism should open a broad avenue for exploring novel
topological quantum matter, and our results have already showcased two prime
examples, i.e., the type-II and type-III TDPs.

Our results may motivate further theoretical and experimental studies of
TDPs and other novel topological matter. Although our proposed experimental
scheme is for cold-atom optical lattices, similar type-II and type-III TDPs may
also be found in some solid-state materials in certain space groups
by first-principles calculations~\cite{Weng} and
angle-resolved photoemission spectroscopy experiments. 
Moreover, we note that recently
a type-I TDP has been experimentally realized in the parameter space 
of a superconducting qutrit~\cite{SQ}, 
where the type-II and type-III TDPs may also be realized similarly. Finally,
although we focus on the spin-$1$ rank-$2$ tensors for the purpose of
studying TDPs, there exist higher-rank spin-tensors for higher spin systems,
whose couplings with momentum may give rise to 
nontrivial topological matter with unprecedented properties.

\begin{acknowledgments}
This work is supported by NSF (PHY-1505496), ARO (W911NF-17-1-0128), AFOSR
(FA9550-16-1-0387), and UTD Research Enhancement Funds.
\end{acknowledgments}

\clearpage
\onecolumngrid
\appendix
\section{Supplementary Materials}
\subsection{Proof of $|\mathcal{C}|\leq 2$ for TDPs of Linear Hamiltonians} 

In this section, we unveil the geometric meaning of the topological
invariant defined in Eq.~(1) of the main text and give an intuitive yet rigorous
proof for $|\mathcal{C}|\leq 2$ in a more general setting. To this end, we
introduce a powerful tool---Majorana stellar representation~\cite{majorana1}, 
which maps quantum states in a high-dimensional Hilbert space onto several
points (i.e., Majorana stars) on the Bloch sphere---the state space of a quantum
spin-1/2 system.  In this representation, any spin-1 state can be mapped to
two Majorana stars on the Bloch sphere. For convenience, the integral
surface $\bm{S}$ in Eq.~(1) is chosen as the unit sphere.

We start with the well-known spin-1/2 system. In a chosen basis (denoted as $%
|\uparrow \rangle $, $|\downarrow \rangle $), an arbitrary state can be
written as $|u\rangle =\cos \frac{\theta }{2}|\uparrow \rangle +e^{i\phi
}\sin \frac{\theta }{2}|\downarrow \rangle $ ($0\leq \theta \leq \pi $, $%
0\leq \phi < 2\pi $). The state $|u\rangle $ is represented by a point $%
\bm{u}=(\sin \theta \cos \phi ,\sin \theta \sin \phi ,\cos \theta )$ on the
Bloch sphere, with $\theta $ and $\phi $ denoting the colatitude and
longitude in the spherical coordinate. For a Weyl point $H(\bm{k})=-\bm{k}%
\cdot \bm{\sigma}$, $|u\rangle $ is the lower state at $\hat{\bm k}=(\sin \theta
\cos \phi ,\sin \theta \sin \phi ,\cos \theta )$, that is, the Majorana star
$\bm{u}$ on the Bloch sphere coincides with $\hat{\bm{k}}$ on the integral surface
$\bm{S}$. The Chern number (monopole charge) of the Weyl point is then
\begin{equation}
\mathcal{C}=\frac{1}{2\pi }\oint_{\bm{S}}\bm{\Omega}(\bm{k})\cdot d\bm{S}=-%
\frac{1}{4\pi }\oint_{\bm{S}}d\theta d\phi ~\bm{u}\cdot \partial _{\theta }%
\bm{u}\times \partial _{\phi }\bm{u}=-1.
\end{equation}%
Clearly, $\mathcal{C}$ counts how many times the Majorana star covers the
Bloch sphere by varying $\hat{\bm k}$ on $\bm S$.

For a spin-1 system, any quantum state can be formulated as $|\psi \rangle
=f_{-1}|1,-1\rangle +f_{0}|1,0\rangle +f_{1}|1,1\rangle $ in a given basis $%
|1,m\rangle ~(m=\pm 1,0)$. The basis state can be rewritten using the
creation and annihilation operators $a^{\dag }$, $a$, and $b^{\dag }$, $b$
of Schwinger bosons~\cite{schwinger}: $|1,m\rangle =\frac{(a^{\dag
})^{1+m}(b^{\dag })^{1-m}}{(1+m)!(1-m)!}|\emptyset \rangle$ ($|\emptyset
\rangle $ is a vacuum state). The Schwinger bosons satisfy the standard  bosonic commutation relations: $[a,a^{\dag}]=[b,b^{\dag}]=1$ and all others are zero. The spin-1 operators are represented by two types of Schwinger bosons as:
\begin{equation}
F^{+}=F_x+i F_y=a^{\dag}b,~~~F^{-}=F_x-i F_y=b^{\dag}a,~~~F_z=\frac{1}{2}(a^{\dag}a-b^{\dag}b),
\end{equation}
along with the constraint $n_a+n_b\equiv a^{\dag}a+b^{\dag}b=2F$. Here $n_a$ and $n_b$ are the occupation numbers of Schwinger bosons. The spin-1 basis state $|1,m\rangle$ is then equivalent to the state $|n_a,n_b\rangle=|1+m,1-m\rangle$. It is easy to verify the commutation relations for spin operators: $[F_i,F_j]=i\epsilon_{ijk}F_k$. Now the spin-1 state $|\psi \rangle $ can be factorized 
as~\cite{majorana1,schwinger,majorana2,majorana3,majorana4,majorana5}
\begin{equation}
|\psi \rangle =\frac{1}{N_{1}}\prod_{j=1}^{2}(\cos \frac{\theta _{j}}{2}%
a^{\dag }+\sin \frac{\theta _{j}}{2}e^{i\phi _{j}}b^{\dag })|\emptyset
\rangle ,  \label{ms}
\end{equation}%
where $N_{1}$ is the normalization factor, and the parameters $\theta
_{j}$ and $\phi _{j}$ can be determined by $\sum_{j=0}^{2}\frac{%
(-1)^{j}f_{1-j}}{\sqrt{(2-j)!j!}}y^{2-j}=0$ with $y_{j}=\tan \frac{\theta
_{j}}{2}e^{i\phi _{j}}$. By denoting $a^{\dag }|\emptyset \rangle =|\uparrow
\rangle $, $b^{\dag }|\emptyset \rangle =|\downarrow \rangle $, 
it follows from Eq.~(\ref{ms}) that $|\psi\rangle $ is represented by the two Majorana stars located
at $\bm{u}_{j}=(\sin \theta _{j}\cos \phi _{j},\sin \theta _{j}\sin \phi
_{j},\cos \theta _{j})$ ($j=1,2$) on the Bloch sphere. Within the Majorana
stellar representation, now we are ready to prove $|\mathcal{C}|\leq 2$ for a
spin-1 TDP.
\begin{figure}[h]
\centering
\includegraphics[width=3.6in]{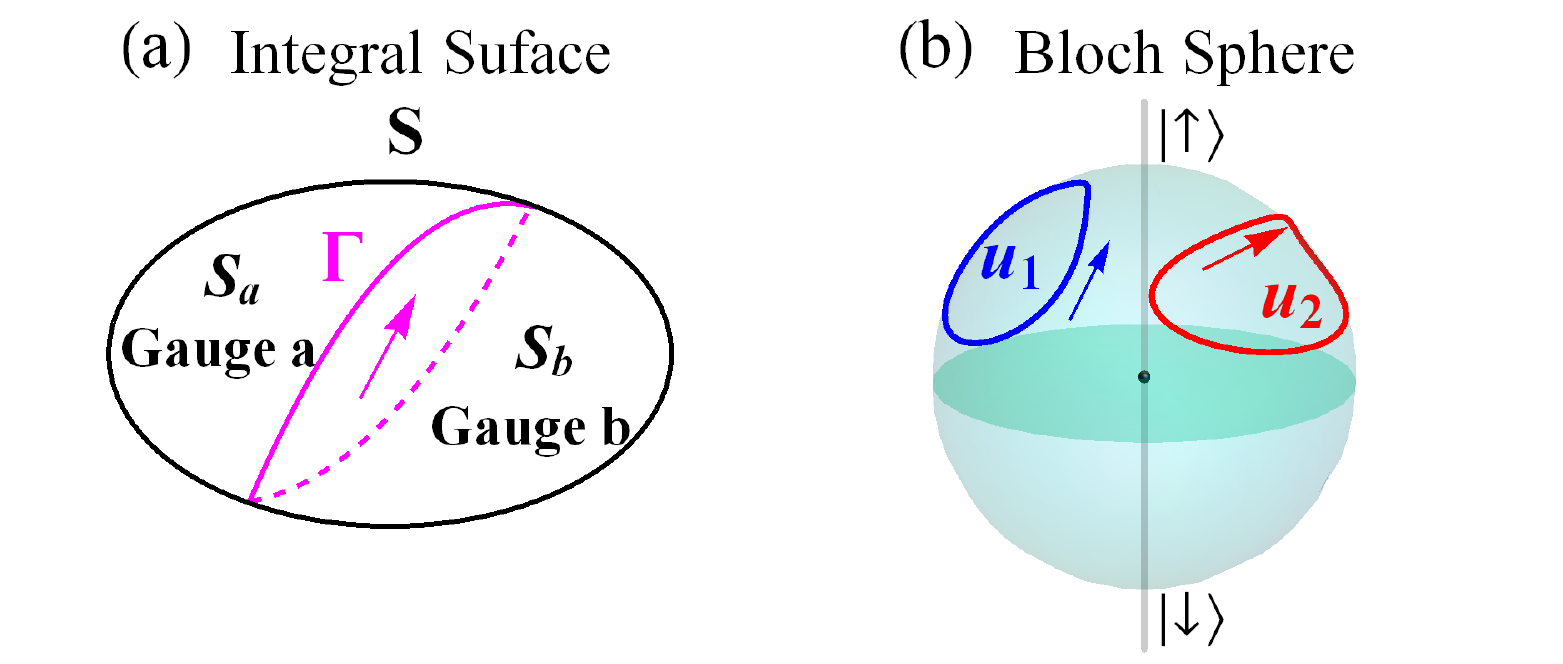}
\caption{(a) The integral surface $\bm{S}$ in momentum space is split into two pieces. In each piece, we can choose a smooth gauge. $\Gamma $
is the boundary between the two pieces. (b) A spin-1 state is
represented by two Majorana stars $\bm{u}_{1}$ and $\bm{u_2}$ on the Bloch
sphere. The Berry phase is determined by the trajectories of two Majorana
stars.}
\label{figchern2}
\end{figure}

Because the Chern number is defined on a closed two-dimensional surface $\bm{S}$
with no boundary, a nonzero Chern number indicates that we cannot choose a
gauge that is continuous and single valued on the whole surface $\bm{S}$ (which yields $\mathcal{C}=0$ by Stokes' theorem). $\bm{S}$ is then separated into
different regions as sketched in Fig.~\ref{figchern2}(a). Inside each
region, we can choose a smooth gauge and use the Stokes' theorem:
\begin{eqnarray}
2\pi \mathcal{C} &=&\oint_{\bm{S}}\bm{\Omega}(\bm{k})\cdot d\bm{S}=\iint_{%
\bm{S}_{a}}{\bm{\Omega}}(\bm{k})\cdot d\bm{S}+\iint_{\bm{S}_{b}}{\bm{\Omega}}%
(\bm{k})\cdot d\bm{S}  \notag \\
&=&\int_{\Gamma }\bm{A}^{a}\cdot d\bm{l}-\int_{\Gamma }\bm{A}^{b}\cdot d%
\bm{l}=\gamma ^{a}-\gamma ^{b}.
\end{eqnarray}%
Here $\bm{A}^{a}$ and $\bm{A}^{b}$ are the gauge potentials associated with
Berry curvature $\bm{\Omega}(\bm{k})$ in each region: $\bm{\nabla}\times %
\bm{A}^{a,b}=\bm{\Omega}(\bm{k})$. $\gamma ^{a}$ and $\gamma ^{b}$ are the
accumulated Berry phases along the path $\Gamma $ (i.e., the boundary of $\bm{S}_{a}$
and $\bm{S}_{b}$) under different gauges. Although $\bm{A}^{a,b}$ and $%
\gamma ^{a,b}$ are gauge-dependent, $\bm{\Omega}(\bm{k})$ is not. From the
Majorana stellar representation, the Berry phase for a spin-1 system in a
chosen gauge can be elegantly formulated as \cite%
{majorana2,majorana3,majorana4}
\begin{equation}
\gamma =\gamma _{S}+\gamma _{C}\equiv -\sum_{j=1}^{2}\frac{1}{2}\oint
(1-\cos \theta _{j})d\phi _{j}-\frac{1}{2}\oint \frac{(d\bm{u}_{1}-d\bm{u}%
_{2})\cdot (\bm{u}_{1}\wedge \bm{u}_{2})}{3+\bm{u}_{1}\cdot \bm{u}_{2}}.
\label{bphase}
\end{equation}%
The first term $\gamma _{S}=\sum_{j=1}^{2}\int_{\Gamma }\langle u_{j}|i%
\bm{\nabla}|u_{j}\rangle \cdot d\bm{l}$ describes the contributions from the
solid angles subtended by the trajectories of two Majorana stars, as shown
in Fig.~\ref{figchern2}(b). While the second term, which is gauge invariant
\cite{majorana5}, comes from their correlations. It is clear from Eq. (\ref%
{bphase}) that a nonzero Chern number solely comes from the gauge mismatch
of the two Majorana stars. Using Stokes' theorem,
\begin{eqnarray}
\mathcal{C} &=&\frac{1}{2\pi }(\gamma _{S}^{a}-\gamma _{S}^{b})=-\frac{1}{%
2\pi }\sum_{j=1}^{2}\text{Im}{\large [\iint_{\bm{S}_{a}}\langle \bm{\nabla}%
u_{j}^{a}|\times |\bm{\nabla}u_{j}^{a}\rangle +\iint_{\bm{S}_{b}}\langle %
\bm{\nabla}u_{j}^{b}|\times |\bm{\nabla}u_{j}^{b}\rangle ]\cdot d{\bm{S}}}
\notag \\
&=&-\frac{1}{2\pi }\sum_{j=1}^{2}\text{Im}\oint_{\bm{S}}(\langle \bm{\nabla}%
u_{j}|\times |\bm{\nabla}u_{j}\rangle )\cdot d{\bm{S}}=-\frac{1}{4\pi }%
\sum_{j=1}^{2}\oint_{\bm{S}}d\theta d\phi ~\bm{u_j}\cdot \partial _{\theta }%
\bm{u_j}\times \partial _{\phi }\bm{u_j}.
\label{C2}
\end{eqnarray}%
Geometrically, $\mathcal{C}$ is the sum of the covering numbers of the two
Majorana stars on the Bloch sphere. To prove $|\mathcal{C}|\leq 2$, we only
need to show, each Majorana star covers Bloch sphere at most once for our
system. In another words, given two Majorana stars $\bm{u}_{1}$ and $\bm{u}%
_{2}$ on the Bloch sphere, we can find at most one $\hat{\bm{k}}$ on $\bm{S}$,
with $\bm{u}_{1}$ and $\bm{u}_{2}$ being the projection of the lowest state of $%
H(\hat{\bm{k}})$ in the Majorana stellar representation.

This is done by reductio ad absurdum. We can construct a unique spin-1 state
$|\psi \rangle $ (up to an irrelevant phase) using $\bm{u}_{1}$ and $\bm{u_2}$. 
Suppose both $\hat{\bm{k}}_{1}$ and $\hat{\bm{k}}_{2}$ satisfy the
condition: $H(\hat{\bm{k}_{1}})|\psi \rangle =e_{1}|\psi \rangle $ and $H(\hat{\bm{k}}%
_{2})|\psi \rangle =e_{2}|\psi \rangle $, with $e_{1}$ and $e_{2}$ the lowest-state energies. 
Because our Hamiltonian is traceless, the sum of
all the three eigenvalues must be $0$. It follows that $e_{1,2}<0$ (which cannot be $0$ due
to the gapped spectrum on $\bm{S}$). From $\hat{\bm{k}_1}$ and $\hat{\bm{k}_2}$, 
we can find a point $\hat{\bm{k}}^*=\frac{e_{2}\hat{\bm{k}}_{1}-e_{1}\hat{\bm{k}}_{2}}{|e_{2}\hat{\bm{k}}_{1}-e_{1}\hat{\bm{k}}_{2}|}$ on $\bm{S}$. The linearity of Hamiltonian yields $%
H(\hat{\bm{k}}^*)|\psi \rangle =\frac{1}{|e_{2}\hat{\bm{k}_{1}}-e_{1}\hat{\bm{k}_{2}}|}[e_{2}H(\hat{\bm{k}_{1}})|\psi
\rangle -e_{1}H(\hat{\bm{k}_{2}})|\psi \rangle] =0$, in contradiction to the
traceless nature of the Hamiltonian. Therefore, there is at most 
one $\hat{\bm{k}}$ on $\bm{S}$ for any two given Majorana stars $\bm{u}_{1}$ and $\bm{u}_{2}$ 
on the Bloch sphere. This concludes that $|\mathcal{C}|\leq 2$ in Eq.~(\ref{C2}).


For a linear Hamiltonian with $H(\mathbf{k})=-H(-\mathbf{k})$, we have $%
C_{+1}=-C_{-1}$ for the upper and lower bands and $C_{0}=0$ for the middle
band. $\left\vert \mathcal{C}_{n}\right\vert \leq 2$ determines that there
are only three types of TDPs, classified by $\mathcal{C}=\pm 2,\pm 1,0$, as
discussed in the main text. We note that in the above proof, only the $\bm k$-linear
and traceless properties of the Hamiltonians are used. Therefore, our classification
of TDPs is quite general and can be used for all spin-vector and spin-tensor
momentum coupling cases, given the fact that all the spin-vectors and spin-tensors are traceless. 
Finally, although we consider the traceless
Hamiltonians in the above proof, any additional spin-independent linear term
such as $\eta k_{z}$ in the Hamiltonian only rotates the eigenspectrum in the
momentum space without changing the eigenstates, and therefore all topological
invariances and topological phase transitions do not change.

\subsection{An extended model for TDPs}

Besides the simple model with one spin-tensor momentum coupling term in the
main text, the above general classification of TDPs also applies to more
complicated models with two spin-tensors coupled to momenta, that is,
\begin{equation}
H(\mathbf{k})=\mathbf{k}\cdot \mathbf{F}+\gamma _{1}k_{y}N_{ij}+\gamma
_{2}k_{z}N_{i^{\prime }j^{\prime }}.  \label{model2}
\end{equation}%
The first term is the standard spin-vector-momentum coupling. Without loss
of generality, the two spin-tensors $N_{ij}$ and $N_{i^{\prime }j^{\prime }}$
are respectively coupled to $k_{y}$ and $k_{z}$. $\gamma _{1}$ and $\gamma _{2}$ are the
coupling strengths. In Table~\ref{table:1}, we have listed all the possible new types of TDPs.
\begin{table}[h!]
\newcommand{\tabincell}[2]{\begin{tabular}{@{}#1@{}}#2\end{tabular}}
\centering
\begin{tabular}{||c| c| c|c|c|c|c||}
 \hline
\diagbox{$N_{ij}$}{$N_{i'j'}$} & $N_{xx}$~~~ & $N_{xy}$~~~& $N_{yy}$~~~ & $N_{xz}$~~~ & $N_{yz}$~~~ & $N_{zz}$~~~ \\  [0.5ex]
 \hline\hline
$N_{xx}$		& $\times$	 &$\times$  &$\times$ &  III  & III & II \\
\hline
$N_{xy}$		& III     &   III & III &  III  &  III & II,III\\
\hline
$N_{yy}$		& II & II & II& II,III&II,III &II,III \\
\hline
$N_{xz}$  & $\times$ &$\times$  &$\times$ &III &III &II \\
\hline
$N_{yz}$  & III & III &III &III &II,III &II,III \\
\hline
$N_{zz}$  & $\times$ &$\times$  &$\times$ & III& III&II \\
\hline
\end{tabular}
\caption{Type-II and type-III TDPs induced by two spin-tensor-momentum 
coupling terms via tuning their strengths $\gamma_1$ and $\gamma_2$. 
``$\times$'' means the corresponding spin-tensor-momentum couplings 
cannot change the type of the original TDP at $\gamma_1=\gamma_2=0$, which is always type-I.}
\label{table:1}
\end{table}

It is clear from Table~\ref{table:1} that all induced TDPs still belong to the three
types, classified by different Chern numbers: $\mathcal{C}=\pm 2$, $\pm 1$, $%
0$. The inclusion of more spin-tensor-momentum couplings can trigger more
topological phase transitions, due to the level crossings induced by these
terms. Similarly, we can discuss these level crossings, Zeeman splittings,
etc. Moreover, we have checked all $6\times 6\times 6=216$ cases with three
spin-tensors coupled into the Hamiltonian. These results are in consistent
with our classification and general discussions.

\subsection{Calculation of the topological invariant $\mathcal{C}$}

For a given Hamiltonian $H(\bm{k})$, we can calculate its three eigenstates $%
\left\vert \psi _{n}\left( \bm{k}\right) \right\rangle $, from which we can
determine the Berry curvature $\mathbf{\Omega }_{n}(\mathbf{k})$. The Chern
number of each band is defined as $\mathcal{C}_{n}=\frac{1}{2\pi }\oint_{\bm{S}}\mathbf{%
\Omega }_{n}(\mathbf{k})\cdot d\mathbf{S}$, where the integral surface $%
\mathbf{S}$ is chosen as a sphere of radius $k$ around the TDP, and the
surface element $d\mathbf{S}=k^{2}\sin \theta d\theta d\phi \frac{\mathbf{k}%
}{k}$.

For the standard Hamiltonian $\bm{k}\cdot \bm{F}$, the
eigenvalues are $-k,0,k$; by taking $\bm{k}=k(\sin \theta \cos \phi ,\sin
\theta \sin \phi ,\cos \theta )$, the corresponding eigenstates are $%
|-1\rangle =\left( \sin ^{2}\frac{\theta }{2}e^{-i\phi },-\frac{\sin \theta
}{\sqrt{2}},\cos ^{2}\frac{\theta }{2}e^{i\phi }\right) ^{T}$, $|0\rangle
=\left( -\frac{\sin \theta }{\sqrt{2}}e^{-i\phi },\cos \theta ,\frac{\sin
\theta }{\sqrt{2}}e^{i\phi }\right) ^{T}$, $|1\rangle =\left( \cos ^{2}\frac{%
\theta }{2}e^{-i\phi },\frac{\sin \theta }{\sqrt{2}},\sin ^{2}\frac{\theta }{%
2}e^{i\phi }\right) ^{T}$. The resulting Berry curvature for each band is
found to be $\mathbf{\Omega }_{n}(\mathbf{k})=-n\mathbf{k}/k^{3}$,
yielding $\mathcal{C}_{n}=\frac{1}{2\pi }\int_{0}^{\pi }k^{2}\sin \theta
d\theta \int_{0}^{2\pi }d\phi (-n\frac{k}{k^{3}})=-n\int_{0}^{\pi }\sin
\theta d\theta =-2n$. For comparison, the Berry curvature of a spin-1/2 is $%
\mathbf{\Omega }_{n}(\mathbf{k})=-n\frac{\mathbf{k}}{2k^{3}}$ \cite{book},
which gives $\mathcal{C}_{n}=-n$. As $n=\mp 1$, $\mathcal{C}_{n}=\pm 1$.

For a general Hamiltonian with spin-tensors, the eigenstates and Berry
curvatures cannot be determined analytically, therefore all calculations are
done numerically.

\subsection{Determination of phase transition points}

The inclusion of spin-tensors $N_{ij}$ can induce a series of topological
phase transitions, accompanied by level-crossings in $\bm k$ space. To
determine these phase transition points and level-crossing lines
analytically, we utilize the traceless property of the Hamiltonian (all the
spin-vectors $F_i$ and spin-tensors $N_{ij}$ are traceless), which dictates that
the sum of the three eigenvalues is zero. For our model (2) with $\alpha\neq
0$, $H(\bm k)=k_x F_x +k_y F_y+\alpha k_z(F_z+\gamma N_{ij})$, here $%
\gamma=\beta/\alpha$. The topological properties would not change by
rescaling $k_z$. For simplicity, we directly set $\alpha=1$ and the integral
surface $\bm{S}$ is chosen as the unit sphere with $k=1$. Suppose two
bands touch at some specific $\bm{k}$, at which the three eigenenergies are given by $E_a$, $E_a$, and $-2E_a$, then
\begin{eqnarray}
\text{det}(xI-H(\bm k))=(x-E_a)(x-E_a)(x+2E_a)=x^3-3E_a^2x+2E_a^3\equiv x^3+d_1x+d_0,
\end{eqnarray}
where $d_1$ and $d_0$ satisfy $P(\bm k)\equiv -d_1^3/27-d_0^2/4=0$. In the following, we determine the phase transition conditions using $P(\mathbf{k})$. If $P(\mathbf{k})$ cannot be zero, then there is no phase transitions as no level crossings are allowed by tuning parameters. For the $%
6$ spin-tensors, we find the following results (by setting $y= k_z^2\gamma^2$).

(A) $N_{xx}$, $N_{yy}$, and $N_{xy}$ would not induce any band crossing. Consider $N_{xx}$ as an example. 
$P(\bm k)=k_x^2y^2/27+(-k_x^4/4+k_x^2/6+1/108)y+1/27$. As $k=1$, $P(\mathbf{k})\geq k_x^2y^2/27+(-k_x^2/4+k_x^2/6+k_x^2/108)y+1/27=k_x^2(y-1)^2/27$. Here ``$=$'' is exact for $|k_x|=1$, hence $y\neq 1$ on the unit sphere $\bm{S}$ and we have $P(\mathbf{k})>0$. Similarly, we have
\begin{eqnarray*}
&&N_{yy}:~P(\bm k)=k_y^2y^2/27+(-k_y^4/4+k_y^2/6+1/108)y+1/27\geq
k_y^2(y-1)^2/27>0; \\
&&N_{xy}:~P(\bm k)=y^3/1728+y^2/144-k_x^2k_y^2 y/4+y/36+1/27\geq
y^3/1728+y^2/144-5y/144+1/27> 0.
\end{eqnarray*}
For all the above three cases, the TDP is still type-I.

(B) For $N_{zz}$, $P(\bm k)=k_z^2y^2/27+(-k_z^4/4+k_z^2/6+1/108)y+1/27\geq
k_z^2(y-1)^2/27\geq 0$. ``$=$'' is valid only when $k_z^2=1$ and $\gamma^2=1$%
, which is the level-crossing point. Specifically, for $\gamma=1$, the lower
(upper) band and middle band touch at $k_z=1 (-1)$; for $\gamma=-1$, the
upper (lower) band and middle band touch at $k_z=1 (-1)$.

(C) For $N_{xz}$, $P(\bm k)=y^3/1728+y^2/144-k_x^2k_z^2 y/4+y/36+1/27\geq
y^3/1728+y^2/144-5y/144+1/27\geq 0$. ``$=$'' is valid when $\gamma=\pm2$ and
$k_x^2=k_z^2=1/2$. At $\gamma=2$, the lower band and middle band touch at $%
\pm k_z=k_x=1/\sqrt{2}$. The upper band and middle band touch at $\pm
k_z=-k_x=1/\sqrt{2}$. Similar analysis can be applied to another transition
point $\gamma=-2$. Note that for $N_{yz}$ the results would be the same, 
by considering $k_y\rightarrow k_x$ and $F_y\rightarrow F_x$.

(D) $\alpha=0$. In this case, $H(\bm k)=k_x F_x+k_y F_y+\beta k_z N_{ij}$. For $N_{xx}$, $N_{xy}$, $N_{yy}$, and $N_{zz}$, there
exist nodal lines where two bands touch in the band structure (the triply-degenerate node is not
the only degenerate point). Thus the Chern number is
ill-defined. For $N_{xz}$ and $N_{yz}$, the eigenenergies of $H(\bm k)$ are given by $0$, and $%
\pm\sqrt{k_x^2+k_y^2+k_z^2\beta^2/4}$. The band structure is adiabatically connected to
the case (C) with $|\gamma|>2$.

As a final remark, the function $P(\bm k)$ can also be used to determine the
splitting of TDPs.

\subsection{Splitting of TDPs by a Zeeman term}

A small Zeeman term $\varepsilon F_{z}$ breaks the triple degeneracy at $%
\bm{k}=0$ for type-I and type-II TDPs described by the Hamiltonian~(2). As a
result, TDPs break into three doubly degenerate Weyl points located at $%
W_{\pm }=(0,0,-\varepsilon /(\alpha \pm \beta ))$ and $W_{3}=(0,0,-%
\varepsilon /\alpha )$. Near these three nodes with $|\delta \mathbf{k}|\ll 1
$, the Hamiltonian reduces to
\begin{eqnarray}
H_{W_{+}}\left( \delta \mathbf{k}\right)  &=&\left( {%
\begin{array}{ccc}
(\alpha +\beta )\delta k_{z} & (\delta k_{x}-i\delta k_{y})/\sqrt{2} & 0 \\
(\delta k_{x}+i\delta k_{y})/\sqrt{2} & 0 & (\delta k_{x}-i\delta k_{y})/\sqrt{2} \\
0 & (\delta k_{x}+i\delta k_{y})/\sqrt{2} &   \frac{2\alpha
\varepsilon }{\alpha +\beta }%
\end{array}%
}\right) , \\
H_{W_{-}}\left( \delta \mathbf{k}\right)  &=&~\left( {%
\begin{array}{ccc}
-\frac{2\beta \varepsilon }{\alpha -\beta } & (\delta k_{x}-i\delta k_{y})/\sqrt{2} & 0
\\
(\delta k_{x}+i\delta k_{y})/\sqrt{2} & 0 & (\delta k_{x}-i\delta k_{y})/\sqrt{2} \\
0 & (\delta k_{x}+i\delta k_{y})/\sqrt{2} & (-\alpha +\beta )\delta k_{z}%
\end{array}%
}\right) , \\
H_{W_{3}}\left( \delta \mathbf{k}\right)  &=&\left( {%
\begin{array}{ccc}
-\frac{\beta \varepsilon }{\alpha }+(\alpha +\beta )\delta k_{z} & (\delta
k_{x}-i\delta k_{y})/\sqrt{2} & 0 \\
(\delta k_{x}+i\delta k_{y})/\sqrt{2} & 0 & (\delta k_{x}-i\delta k_{y})/\sqrt{2} \\
0 & (\delta k_{x}+i\delta k_{y})/\sqrt{2} & -\frac{\beta \varepsilon }{\alpha }%
+(-\alpha +\beta )\delta k_{z}%
\end{array}%
}\right) .
\end{eqnarray}%
Therefore the effective two-band Hamiltonians can be expressed as
\begin{eqnarray}
H_{W_{+}}(\delta \bm{k}) &=&\frac{1}{\sqrt{2}}\delta k_{x}\sigma _{x}+\frac{1%
}{\sqrt{2}}\delta k_{y}\sigma _{y}+\frac{\alpha +\beta }{2}\delta
k_{z}\sigma _{z}+\frac{\alpha +\beta }{2}\delta k_{z}I_{2}+O(\delta k^{2}),
\\
H_{W_{-}}(\delta \bm{k}) &=&\frac{1}{\sqrt{2}}\delta k_{x}\sigma _{x}+\frac{1%
}{\sqrt{2}}\delta k_{y}\sigma _{y}+\frac{\alpha -\beta }{2}\delta
k_{z}\sigma _{z}+\frac{\beta -\alpha }{2}\delta k_{z}I_{2}+O(\delta k^{2}),
\end{eqnarray}%
up to the linear order of $\delta \bm{k}$ and
\begin{eqnarray}
H_{W_{3}}(\delta \bm{k}) &=&\alpha \delta k_{z}\sigma _{z}-\frac{\alpha}{2\beta\varepsilon}[(\delta k_x^2-\delta k_y^2)\sigma_x+2\delta k_x\delta k_y\sigma_y] +(\beta \delta
k_{z}-\frac{\beta \varepsilon }{\alpha })I_{2}+ O(\delta k^3),
\end{eqnarray}
 up to second order of $\delta \bm{k}$. The first two are Weyl points
with linear dispersions along all three directions, whereas the third one is
a multi-Weyl point which has a linear dispersion in the $k_{z}$ direction
but quadratic dispersion along the other two directions. The Chern numbers
for this multi-Weyl point is $\mathcal{C}=2$. The quadratic dispersion
originates from the non-direct (second-order) couplings in $F_{x}$ and $F_{y}$
between the degenerate energy levels ($|+1\rangle $ and $|-1\rangle $). 
For $|\beta |<|\alpha |$, the
linear Weyl points $W_{\pm }$ have the same charge $\mathcal{C}=1$ ($\alpha
>0$), i.e., the case for type-I TDPs. For $|\beta |>|\alpha |$, the linear
Weyl points $W_{\pm }$ have opposite charges $\mathcal{C}=\pm 1$, i.e., the
case for type-II TDPs.

In the lattice model described by Eq.~(3), two TDPs appear at $(0,0,\pm \arccos
(-\gamma ))$. By adding a Zeeman term $\varepsilon F_{z}$, the TDP at $%
(0,0,\arccos (-\gamma ))$ is split into three nodes at $k_{1}=\arccos [-%
\frac{\varepsilon }{t_{0}(1+\beta )}-\gamma ]$, $k_{2}=\arccos [-\frac{%
\varepsilon }{t_{0}(1-\beta )}-\gamma ]$, and $k_{3}=\arccos (-\frac{%
\varepsilon }{t_{0}}-\gamma )$ along the $k_{x}=k_{y}=0$ line. Around the
first two degenerate nodes, the effective two-band Hamiltonians can be
written as
\begin{equation}
H_{k_{1,2}}=\frac{1}{\sqrt{2}}(\delta k_{x}\sigma _{x}+\delta k_{y}\sigma
_{y})-\frac{t_{0}\sin k_{1,2}}{2}(1\pm \beta )\delta k_{z}\sigma
_{z}+O(\delta k^{2}),  \label{weylham}
\end{equation}%
which describe two linear Weyl points. For $\gamma =-0.5$, both Weyl points
have $\mathcal{C}=-1$ for $0<\beta <1$ (type-I) and $\mathcal{C}=\pm 1$ for $%
\beta >1$ (type-II), which are consistent with our numerical results. The
third multi-Weyl point has $\mathcal{C}=-2$ and can be described by $%
H_{k_{3}}=-t_{0}\sin k_{3}\delta k_{z}\sigma _{z}+O(\delta k^{2})$, whose
energy dispersion is linear in the $k_{z}$ direction and quadratic in the
other two directions. A similar analysis can be applied to the other TDP.

Under the same perturbation, a type-III TDP is broken into four linear Weyl
points located at $(k_{x},k_{z})=(\pm \beta \varepsilon /(\beta -2\alpha
),2\varepsilon /(\beta -2\alpha )),(\pm \beta \varepsilon /(\beta +2\alpha
),-2\varepsilon /(\beta+2\alpha ))$ in the $k_{y}=0$ plane. By neglecting
those constant terms, the effective two-band Hamiltonians around these Weyl
points are given by
\begin{eqnarray}
H_{1}(\delta \bm{k}) &=&\frac{1}{\sqrt{3}}(\delta k_{x}-\frac{\beta }{2}%
\delta k_{z})\sigma _{x}+\frac{1}{\sqrt{3}}\delta k_{y}\sigma _{y}-[\frac{1}{%
3}\delta k_{x}+(\frac{\beta }{6}-\frac{2\alpha }{3})\delta k_{z}]\sigma
_{z}+O(\delta k^{2}), \\
H_{2}(\delta \bm{k}) &=&\frac{1}{\sqrt{3}}(-\delta k_{x}-\frac{\beta }{2}%
\delta k_{z})\sigma _{x}-\frac{1}{\sqrt{3}}\delta k_{y}\sigma _{y}+[\frac{1}{%
3}\delta k_{x}-(\frac{\beta }{6}-\frac{2\alpha }{3})\delta k_{z}]\sigma
_{z}+O(\delta k^{2}), \\
H_{3}(\delta \bm{k}) &=&\frac{1}{\sqrt{3}}(\delta k_{x}+\frac{\beta }{2}%
\delta k_{z})\sigma _{x}+\frac{1}{\sqrt{3}}\delta k_{y}\sigma _{y}-[\frac{1}{%
3}\delta k_{x}-(\frac{\beta }{6}+\frac{2\alpha }{3})\delta k_{z}]\sigma
_{z}+O(\delta k^{2}), \\
H_{4}(\delta \bm{k}) &=&\frac{1}{\sqrt{3}}(\delta k_{x}-\frac{\beta }{2}%
\delta k_{z})\sigma _{x}-\frac{1}{\sqrt{3}}\delta k_{y}\sigma _{y}-[\frac{1}{%
3}\delta k_{x}+(\frac{\beta }{6}+\frac{2\alpha }{3})\delta k_{z}]\sigma
_{z}+O(\delta k^{2}).
\end{eqnarray}%
These four nodal points can be regarded as deformed Weyl points rotated by a
spin-tensor $N_{xz}$ in the $k_y=0$ plane.

Although not in the standard form, the four Weyl points are still
characterized by the Chern numbers defined in Eq.~(1). In principle, the
topological invariants can be determined numerically, as we have done. Here,
we show that several symmetry arguments can be used for determining their
Chern numbers relatively. As the first two Weyl points are related by $%
H_{1}(\delta k_{x},\delta k_{y},\delta k_{z})=H_{2}(-\delta k_{x},-\delta
k_{y},\delta k_{z})$, they must have the same Chern number. As the last two
Weyl points are related by $H_{3}(\delta k_{x},\delta k_{y},\delta
k_{z})=H_{4}(\delta k_{x},-\delta k_{y},-\delta k_{z})$, they must have the
same Chern number, too. Note that the four Weyl points always exist even at $%
\alpha =0$ for a finite Zeeman splitting. By tuning $\alpha $ to $0$, they
move in the $k_{y}=0$ plane without merging. The entire process is adiabatic
because no level touching or crossing occurs.

At $\alpha=0$, as the first and third Weyl points are related by $%
H_{1}(\delta k_{x},\delta k_{y},\delta k_{z})=H_{3}(\delta k_{x},\delta
k_{y},-\delta k_{z})$, and the first and fourth Weyl points are related by $%
H_{1}(\delta k_{x},\delta k_{y},\delta k_{z})=H_{4}(\delta k_{x},-\delta
k_{y},\delta k_{z})$, the first two and the last two Weyl points must have
opposite Chern numbers. Therefore, a type-III TDP can be split into two
pairs of Weyl points with opposite charges, as verified by our numerical
results.

\subsection{Experimental scheme}

\begin{figure}[h]
\centering
\includegraphics[width=3.8in]{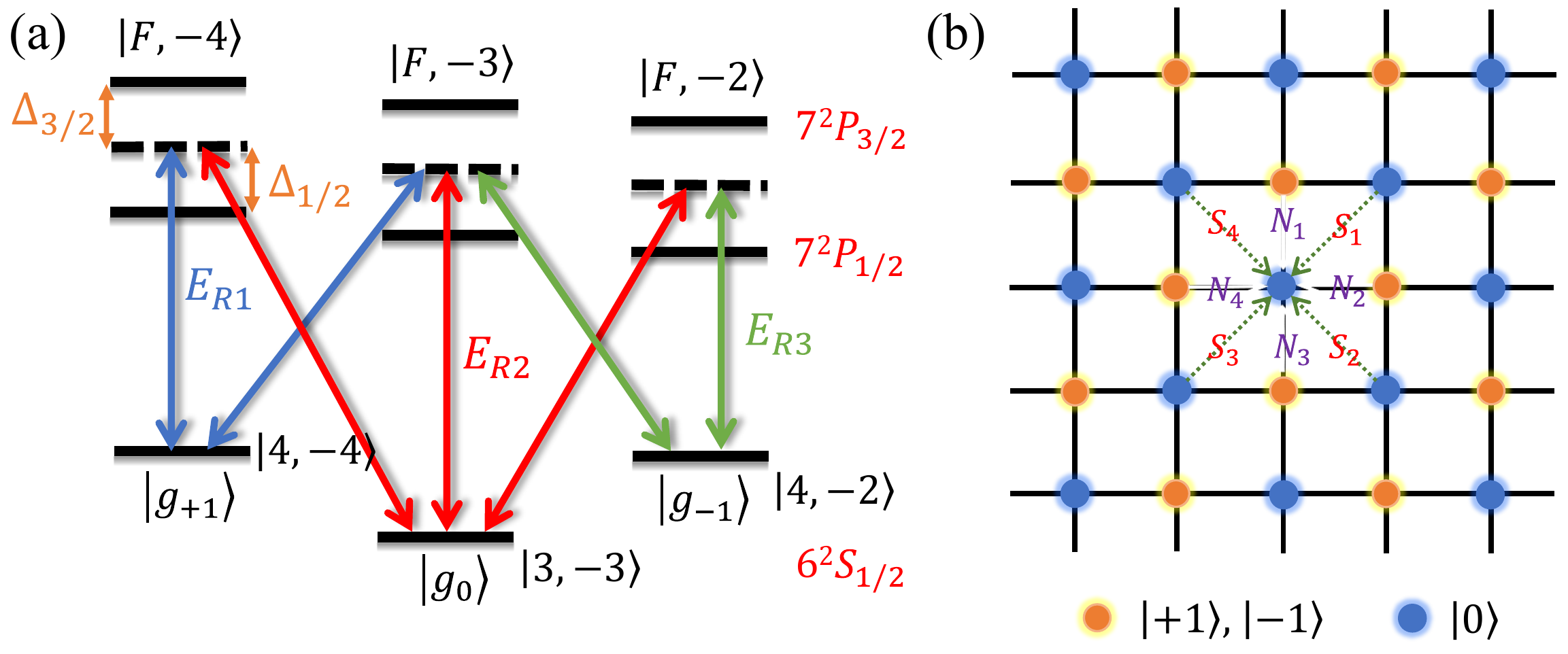}
\caption{(a) Optical transitions to generate Raman couplings between three
hyperfine states. (b) Schematic of the tight-binding model, in which $g_{\pm
1}$ stay in one sublattice while $g_0$ in the other sublattice, ${\bm{N}}%
_{1} $ to ${\bm{N}}_{4}$ denote the nearest-neighbor bonding between
different components, and ${\bm{S}}_{1}$ to ${\bm{S}}_{4}$ denote the
next-nearest-neighbor bonding between the same components.}
\label{raman}
\end{figure}

Here we discuss how to experimentally realize spin-vector- and spin-tensor-momentum 
couplings, which are crucial for engineering different types of
TDPs. Consider the following three Raman beams
\begin{equation*}
\bm{E}_{R_{1},R_{3}}=E_{R_{1},R_{3}}e^{\mp ik_{m}z}[\hat{\bm{x}}\cos
(2k_{0}y)\mp \hat{\bm{y}}\cos (2k_{0}x)]\,,\quad \bm{E}%
_{R_{2}}=E_{R_{2}}e^{ik_{1}z}(i\hat{\bm{x}}+\hat{\bm{y}}).
\end{equation*}%
The $\bm{E}_{R_{1}}$ and $\bm{E}_{R_{3}}$ fields can be formed by multiple
reflections of a beam in a 3D space that is initially polarized along $\hat{%
\bm{x}}$ and incident in the $y$-$z$ plane with an incident angle determined
by $k_{1}^{2}=k_{m}^{2}+4k_{0}^{2}$. The $\bm{E}_{R_{2}}$ beam is a
traveling wave in the $z$ direction with a wavevector $k_{1}$. A magnetic
field $\bm{B}$ is applied in the $x$-$y$ plane with a $\pi /4$-angle with
respect to $\hat{\bm{x}}$. The Raman couplings between the hyperfine states
are contributed from both $D_{1}(6^{2}S_{1/2}\rightarrow 7^{2}P_{1/2})$ and $%
D_{2}(6^{2}S_{1/2}\rightarrow 7^{2}P_{3/2})$ lines with detunings $\Delta
_{1/2}$ and $\Delta _{3/2}$, respectively. The detunings are much larger
than the hyperfine structure. The resulting Raman couplings can be obtained
by summing over all the transitions allowed by the selection rules. For the
purpose of calculations, we need to decompose the electric field as follows:
\begin{eqnarray}
\bm{E}_{R_{1},R_{3}} &=&\frac{E_{R_{1},R_{3}}e^{\mp ik_{m}z}}{\sqrt{2}}%
\{[\cos (2k_{0}y)\mp \cos (2k_{0}x)]\hat{\bm{B}}_{\parallel }-[\cos
(2k_{0}y)\pm \cos (2k_{0}x)]\hat{\bm{B}}_{\perp }\},  \notag \\
\bm{E}_{R_{2}} &=&\frac{E_{R_{2}}e^{ik_{1}z}}{\sqrt{2}}[(1+i)\hat{\bm{B}}%
_{\parallel }+(1-i)\hat{\bm{B}}_{\perp }].
\end{eqnarray}%
The component parallel to (perpendicular to) $\bm{B}$ is used to induce the $%
\pi $ ($\sigma $) transition, as illustrated in Fig.~\ref{raman}.

The Raman coupling between $g_{+1}$ and $g_{0}$ comes from the following two
parts by summing over all possible $F$:
\begin{eqnarray*}
M_{+1,0}^{1} &=&\sum_{J=\frac{1}{2},\frac{3}{2}}^{F}\frac{\Omega
_{g_{+1},F,1\parallel }^{J\ast }\Omega _{g_{0},F,2-}^{J}}{\Delta _{J}}=\frac{%
\sqrt{7}E_{R_{1}}E_{R_{2}}\alpha _{D_{1}}^{2}}{12\sqrt{2}}(\frac{1}{\Delta
_{3/2}}-\frac{1}{\Delta _{1/2}})(1-i)e^{i(k_{1}+k_{m})z}[\cos (2k_{0}x)-\cos
(2k_{0}y)], \\
M_{+1,0}^{2} &=&\sum_{J=\frac{1}{2},\frac{3}{2}}^{F}\frac{\Omega
_{g_{+1},F,1+}^{J\ast }\Omega _{g_{0},F,2\parallel }^{J}}{\Delta _{J}}=\frac{%
\sqrt{7}E_{R_{1}}E_{R_{2}}\alpha _{D_{1}}^{2}}{12\sqrt{2}}(\frac{1}{\Delta
_{3/2}}-\frac{1}{\Delta _{1/2}})(1+i)e^{i(k_{1}+k_{m})z}[\cos (2k_{0}x)+\cos
(2k_{0}y)].
\end{eqnarray*}%
Here $\Omega _{g_{s},F,\parallel}^{J}=e\langle g_{s}|z|F,0,J\rangle \hat{%
\bm{e}}_{z}\cdot \bm{E}$ and $\Omega _{g_{s},F,\pm }^{J}=e\langle
g_{s}|e^{\pm }|F,\pm 1,J\rangle\hat{\bm{e}}_{\pm }\cdot\bm{E}$ are the
transition matrix elements in the basis of the circularly polarized light in
the plane perpendicular to $\bm{B}$.

Similarly, the Raman coupling between $g_{-1}$ and $g_{0}$ can be written as
\begin{eqnarray}
M_{-1,0}^{1} &=&\sum_{J=\frac{1}{2},\frac{3}{2}}^{F}\frac{\Omega
_{g_{-1},F,3\parallel }^{J\ast }\Omega _{g_{0},F,2+}^{J}}{\Delta _{J}}=\frac{%
E_{R_{2}}E_{R_{3}}\alpha _{D_{1}}^{2}}{24\sqrt{2}}(\frac{1}{\Delta _{3/2}}-%
\frac{1}{\Delta _{1/2}})(1-i)e^{i(k_{1}-k_{m})z}[\cos (2k_{0}x)+\cos
(2k_{0}y)], \\
M_{-1,0}^{2} &=&\sum_{J=\frac{1}{2},\frac{3}{2}}^{F}\frac{\Omega
_{g_{-1},F,3-}^{J\ast }\Omega _{g_{0},F,2\parallel }^{J}}{\Delta _{J}}=\frac{%
E_{R_{2}}E_{R_{3}}\alpha _{D_{1}}^{2}}{24\sqrt{2}}(\frac{1}{\Delta _{3/2}}-%
\frac{1}{\Delta _{1/2}})(1+i)e^{i(k_{1}-k_{m})z}[\cos (2k_{0}x)-\cos
(2k_{0}y)].
\end{eqnarray}

If follows that the total Raman couplings between $g_{\pm 1}$ and $g_{0}$
are respectively
\begin{eqnarray}
M_{+1,0} &=&M_{+1,0}^{1}+M_{+1,0}^{2}=M_{0}e^{i(k_{1}+k_{m})z}[\cos
(2k_{0}x)+i\cos (2k_{0}y)], \\
M_{-1,0} &=&M_{-1,0}^{1}+M_{-1,0}^{2}=M_{0}^{\prime
}e^{i(k_{1}-k_{m})z}[\cos (2k_{0}x)-i\cos (2k_{0}y)],
\end{eqnarray}%
where $M_{0}=\frac{\sqrt{7}\alpha _{D_{1}}^{2}E_{R_{1}}E_{R_{2}}}{6\sqrt{2}%
\Delta _{2}}$, $M_{0}^{\prime }=\frac{\alpha _{D_{1}}^{2}E_{R_{2}}E_{R_{3}}}{%
12\sqrt{2}\Delta _{2}}$, and $\frac{1}{\Delta _{2}}=\frac{1}{\Delta _{3/2}}-%
\frac{1}{\Delta _{1/2}}$.

To remove the spatially dependent phase factor in the Raman coupling, we can
use the gauge transformation $U=e^{i(k_{1}F_{z}^{2}+k_{m}F_{z})z}$, which
would not affect other terms. In the rotated frame, the Raman coupling then
becomes
\begin{equation}
H_{R}=\lambda k_{z}(k_{1}F_{z}^{2}+k_{m}F_{z})+[\cos (2k_{0}x)+i\cos
(2k_{0}y)](M_{0}|g_{+1}\rangle \langle g_{0}|+M_{0}^{\prime }|g_{0}\rangle
\langle g_{-1}|)+\mbox{h.c.}
\end{equation}%
with $\lambda =\hbar ^{2}/m$ by neglecting those constant term. Since the
spin-dependent lattice potentials have the same sign for $g_{+1}$ and $g_{-1}
$ components, we can write the tight-binding model on a square lattice in
the $x$-$y$ plane as shown in Fig. \ref{raman}(b), in which $g_{\pm 1}$ stay
in one sublattice while $g_{0}$ in the other sublattice. We consider the
nearest-neighbor and next-nearest-neighbor hopping terms with only $s$%
-orbital of each site. The hopping between the nearest-neighbor sites are
between different components induced by the Raman couplings. The hopping
between the next-nearest-neighbor sites are between the same component. The
effective tight-binding Hamiltonian reads
\begin{eqnarray}
H_{tb} &=&\frac{\lambda k_{z}^{2}}{2}+H_{R}-\sum_{i,j}^{s=\pm
1,0}t_{s}c_{s}^{\dag }(\bm{r_{i}})c_{s}(\bm{r_{i}}+\bm{S}_{j})-\sum_{i}^{s=%
\pm 1,0}\delta _{s}c_{s}^{\dag }(\bm{r_{i}})c_{s}(\bm{r_{i}}) \\
&&+\sum_{i,j}t_{so1}^{ij}c_{+1}^{\dag }(\bm{r}_{i})c_{0}(\bm{r}_{i}+\bm{N}%
_{j})+\sum_{i,j}t_{so2}^{ij}c_{-1}^{\dag }(\bm{r}_{i})c_{0}(\bm{r}_{i}+\bm{N}%
_{j})+\mbox{h.c.},
\end{eqnarray}%
where the Zeeman term has been incorporated into the detunings in the ground
state manifold. The coupling coefficients are
\begin{equation}
t_{s}=\int d^{2}\bm{r}\phi _{s}^{i\ast }\left[ \frac{\lambda }{2}%
(k_{x}^{2}+k_{y}^{2})+V(\bm{r})\right] \phi _{s}^{j}(\bm{r}%
),~~t_{so1}^{ij}=\int d^{2}\bm{r}\phi _{+1}^{i\ast }M_{+1,0}\phi _{0}^{j}(%
\bm{r}),~~t_{so2}^{ij}=\int d^{2}\bm{r}\phi _{-1}^{i\ast }M_{-1,0}\phi
_{0}^{j}(\bm{r}).
\end{equation}%
The spin-flipped hopping coefficients satisfy $t_{so1}^{jx,jx\pm 1}=\pm
t_{so1}$, $t_{so1}^{jy,jy\pm 1}=\pm it_{so1}$, $t_{so2}^{jx,jx\pm 1}=\pm
t_{so2}$, and $t_{so2}^{jy,jy\pm 1}=\mp it_{so2}$, as constrained by the
lattice symmetry. For the spin-dependent lattice, each unit cell contains
two lattice sites with primitive vectors along the two diagonal directions
(lattice constant $b=\pi /k_{0}$). Using Fourier transformation and setting $%
t_{so1}=t_{so2}=\frac{t_{so}}{2\sqrt{2}}$, which can be achieved by
adjusting the relative strengths of Raman beams, we obtain the following
momentum-space Hamiltonian
\begin{equation}
H_{3D}(\bm{k})=\frac{\lambda k_{z}^{2}}{2}-4T_{s}\cos (k_{x}a)\cos
(k_{y}a)-\Lambda _{s}+\lambda
k_{z}(k_{1}F_{z}^{2}+k_{m}F_{z})+t_{so}F_{x}\sin (k_{x}a)+t_{so}F_{y}\sin
(k_{y}a).
\end{equation}%
Here $a=\frac{\pi }{\sqrt{2}k_{0}}$, and $k_{x}=(k_{+}+k_{-})/\sqrt{2}$, $%
k_{y}=(k_{+}-k_{-})/\sqrt{2}$ are lattice momenta along $x$ and $y$
directions. $T_{s}=\mbox{diag}\left( t_{+1},t_{0},t_{-1}\right) $ and $%
\Lambda _{s}=\mbox{diag}\left( \delta _{+1},\delta _{0},\delta _{-1}\right) $
are diagonal matrices for tunneling and detuning. When $t_{+1}=t_{0}=t_{-1}$
and $\delta _{+1}=\delta _{0}=\delta _{-1}$, i.e., no Zeeman term, there
exist two TDPs in the 2D Brillouin zone spanned by $(k_{x},k_{y})$. They are
located at $(0,0)$ and $(\pi ,0)$. (Note that $(0,0)$ and $(\pi ,\pi )$ ( $%
(\pi ,0)$ and $(0,\pi )$) are the same momenta by folding back to the first 
Brillouin zone spanned by $(k_{+},k_{-})$). By expanding the above
Hamiltonian around the two points, we obtain the following low-energy
Hamiltonians (setting $a=1$)
\begin{eqnarray}
H_{1}(\delta \bm{k}) &=&\lambda \delta
k_{z}(k_{1}F_{z}^{2}+k_{m}F_{z})+t_{so}\delta k_{x}F_{x}+t_{so}\delta
k_{y}F_{y}, \\
H_{2}(\delta \bm{k}) &=&\lambda \delta
k_{z}(k_{1}F_{z}^{2}+k_{m}F_{z})-t_{so}\delta k_{x}F_{x}+t_{so}\delta
k_{y}F_{y},  \label{hamlowenergy}
\end{eqnarray}%
which are similar to the Hamiltonian~(2). The two TDPs have the opposite
Chern numbers. When $t_{s}$ are not equal, the resulting Zeeman field at the
two points may be compensated by choosing suitable detuning $\delta _{s}$.
In this case, one of two TDPs will survive, whereas the other one will be
broken into two Weyl points with opposite Chern numbers.
\end{document}